\documentclass[letter,twocolumn]{jpsj2}

\usepackage{times}
\usepackage{graphicx}

\title{Effect of Rattling Phonons on Sommerfeld Constant}

\author{Takashi {\sc Hotta}}

\inst{Department of Physics, Tokyo Metropolitan University,
1-1 Minami-Osawa, Hachioji, Tokyo 192-0397}

\recdate{\today}

\abst{
By employing a numerical renormalization group technique,
we evaluate electronic specific heat coefficient $\gamma$
of the Anderson model coupled with local anharmonic phonons
for the oscillation of a caged atom.
For the rattling-type cage potential
with a flat and wide region in the bottom,
we find that phonon-mediated attraction is largely enhanced.
When the potential shape is deformed from the rattling type,
there occurs a cancellation between Coulomb repulsion and
the phonon-mediated attraction.
In such a situation, spin and charge fluctuations are comparable
to each other,
leading to the realization of exotic electron-phonon complex state
with large and magnetically robust $\gamma$.
}

\kword{Kondo effect, Filled skutterudites, Rattling, Sommerfeld constant}

\begin{document}
\maketitle


Recently, a new possibility of electron-phonon coupled state
emerging in strongly correlated electron materials
such as clathrate compounds,\cite{Takabatake}
pyrochlore oxides,\cite{NMR,Ueda} and
filled skutterudites~\cite{Skut}
has attracted much attention
in the research area of condensed matter physics.
A common feature of these materials is the existence of
nano-size cage composed of relatively light atoms,
in which another atom feels a highly anharmonic potential
and oscillates with large amplitude.
Such an oscillation is frequently called ``rattling'',
which is considered to be one of key ingredients of
cage-structure materials, when we attempt to clarify
their electronic properties.

For instance, in order to understand
magnetically robust heavy-fermion behavior observed in
Sm-based filled skutterudite compound SmOs$_4$Sb$_{12}$,
\cite{Sanada}
the non-magnetic Kondo effect originating from
phonon degree of freedom has been pointed out.
\cite{Miyake,Hattori1,Hattori2}
Along with this research direction,
in order to promote our understandings
on the Kondo physics in electron-phonon systems,
the present author
has performed numerical calculations of
the Anderson model coupled with local Jahn-Teller and
Holstein phonons.\cite{Hotta1,Hotta2,Hotta3}
The periodic Anderson-Holstein model has been also analyzed
with the use of a dynamical mean-field theory and
a mechanism of the mass enhancement due to large lattice fluctuations
and phonon softening towards double-well potential
has been addressed.\cite{Ono}
This tendency has been also suggested in the Holstein model.
\cite{Freericks1,Freericks2,Meyer}

Concerning the robustness of electronic specific heat coefficient $\gamma$
against an applied magnetic field,
it is basically understood, if the Kondo temperature $T_{\rm K}$
becomes large in comparison with the magnetic field.
In fact, in an electron-phonon system, Coulomb interaction
is considered to be, more or less, weakened by
phonon-mediated attraction,
leading to the enhancement of $T_{\rm K}$.
However, in order to cancel the on-site Coulomb interaction,
phonon energy should be in the order of electron bandwidth,
as long as we consider harmonic phonons.
Such a situation may be realized in some molecular conductors
composed of light atoms, but in general, phonon energy
is smaller than the electron bandwidth.
Moreover, when we recall a relation of $\gamma \sim 1/T_{\rm K}$,
\cite{review}
$\gamma$ cannot be so large for the case with enhanced $T_{\rm K}$.

In this paper, in order to examine the phonon-based scenario
for large and magnetically robust $\gamma$,
we investigate the effect of anharmonic oscillation of
the atom in the cage potential by treating them
as local phonons in the standard quantum mechanics.
We analyze the Anderson model coupled with local
anharmonic phonons with the use of
a numerical renormalization group method.
For a rattling-type potential in which
a flat and wide region appears in the bottom,
the Coulomb repulsion is largely suppressed
by phonon-mediated attraction, even when we consider
phonon energy smaller than the electron bandwidth.
We also find that $\gamma$ is both large and
magnetically robust,
when the potential shape is deformed from the rattling type
in which spin fluctuations are comparable to charge ones.


First let us discuss the local electron-phonon state,
when we include anharmonicity.
The local Hamiltonian is given by
\begin{equation}
  H_{\rm loc}=U n_{\uparrow}n_{\downarrow}+H_{\rm eph},
\end{equation}
where $n_{\sigma}$=$d^{\dag}_{\sigma}d_{\sigma}$,
$d_{\sigma}$ is an annihilation operator of localized electron
with spin $\sigma$, $U$ is the Coulomb interaction,
and $H_{\rm eph}$=$g Q \rho$+$P^2/2$+$V(Q)$.
Here $g$ is the electron-phonon coupling constant,
$\rho$=$n_{\uparrow}$+$n_{\downarrow}$,
$Q$ is normal coordinate of the oscillation of a caged atom,
$P$ is the corresponding canonical momentum,
and $V(Q)$ indicates the cage potential.
Note that the reduced mass of the vibration is set as unity.
The potential is given by
$V(Q)$=$\omega^2Q^2/2$+$k_4Q^4$+$k_6Q^6$,
where $\omega$ is the phonon energy
and we consider fourth- and sixth-order
anharmonicity in the potential.
Using the phonon operator $a$ defined through
$Q$=$(a+a^{\dag})/\sqrt{2\omega}$, we obtain
\begin{equation}
  \begin{split}
   H_{\rm eph} &=
   \omega\sqrt{\alpha}(a+a^{\dag}) \rho + \omega(a^{\dag}a+1/2) \\
   &+ \beta_4 \omega (a+a^{\dag})^4+\beta_6 \omega (a+a^{\dag})^6,
  \end{split}
\end{equation}
where the non-dimensional electron-phonon coupling
constant $\alpha$ is given by $\alpha$=$g^2/(2\omega^3)$
and non-dimensional anharmonicity parameters $\beta_4$ and
$\beta_6$ are, respectively, given by
$\beta_4$=$k_4/(4\omega^3)$ and $\beta_6$=$k_6/(8\omega^4)$.
By using these non-dimensional parameters,
we express the potential as
\begin{equation}
  V(q)=\alpha\omega(q^2+16\alpha\beta_4 q^4 + 64 \alpha^2\beta_6 q^6),
\end{equation}
where $q$ is the non-dimensional length, given by $q$=$Q\omega^2/g$.

In this paper, we fix $\beta_6$ as $\beta_6$=$10^{-5}$
and change $\beta_4$ in the negative side in order to
examine the effect of potential shape.
Note that for $\beta_4$$>$0, there is no drastic change
in the potential shape.
As for $\alpha$, we fix $\alpha$=2 to consider
the strong electron-phonon coupling.
In the calculations, we define the phonon basis as
$|\ell \rangle$=$(a^{\dag})^{\ell}|0\rangle/\sqrt{\ell!}$,
where $\ell$ is the phonon number and
$|0\rangle$ is the vacuum state.
The phonon basis is truncated at a finite number,
which is set as 2000.

\begin{figure}[t]
\label{fig1}
\centering
\includegraphics[width=6truecm]{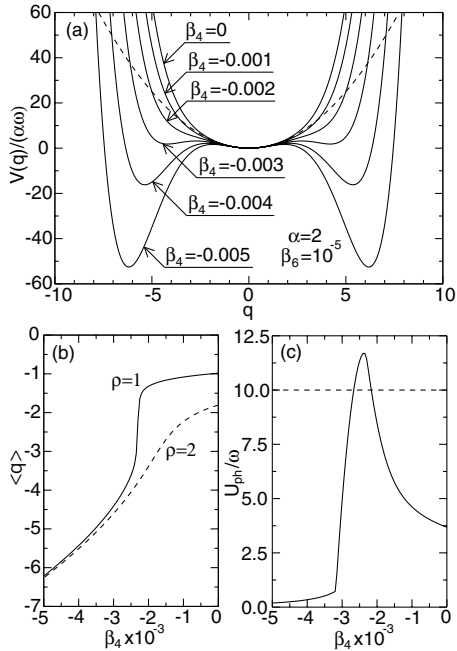}
\caption{
(a) Change of the potential shape due to $\beta_4$
for $\alpha$=2 and $\beta_6$=$10^{-5}$.
Broken curve denotes the quadratic potential.
(b) Average value of $q$ vs. $\beta_4$ for $\rho$=1 and 2.
(c) Phonon-mediated attraction $U_{\rm ph}/\omega$ vs. $\beta_4$.
}
\end{figure}


In Fig.~1(a), we depict $V(q)/(\alpha\omega)$ in order to
see how the potential shape is changed
when $\beta_4$ is decreased in the negative side.
For $\beta_4$=0, there is a single minimum at $q$=0.
In comparison with the case of quadratic potential,
the anharmonic potential becomes steep
for large $q$, while there is no significant difference
among the potentials in the region near $q$=0.
Thus, the potential in this parameter region is called
the on-center type.
When $\beta_4$ is decreased,
shoulder-like structure gradually appears and
the potential bottom becomes flat and wide.
In the present definitions of parameters,
potential minima at $|q|$$\ne$0 appear
for $\beta_4$$<$$-\sqrt{3\beta_6/4}$=$-0.00274$.
This shape of the potential is called the rattling type,
since the caged atom is expected to oscillate
with large amplitude in spite of the structure
with shallow minima in the wide bottom of the potential.
For further decrease of $\beta_4$
with $\beta_4$$<$$-\sqrt{\beta_6}$=$-0.00316$,
the minimum values of
$V(q)$ at $q$$\ne$0 are smaller than that at the origin.
The shape of the potential in this region is called
the off-center type in the sense that the atom
oscillates around at off-center positions due to deep
potential minima at $|q|$$\ne$0.

Here we summarize the three types of the potential shape:
(i) The on-center type
for $-\sqrt{3\beta_6/4}$$<$$\beta_4$$<$0,
(ii) the rattling type
for $-\sqrt{\beta_6}$$<$$\beta_4$$<$$-\sqrt{3\beta_6/4}$,
and (iii) the off-center type
for $\beta_4$$<$$-\sqrt{\beta_6}$.
When we include the electron-phonon coupling,
the boundary values will be changed,
but the above three types still characterize
the potential shape.


In order to see the effect of the potential shape on the
electron state, let us evaluate the average displacement
$\langle q \rangle$ for $\rho$=1 and 2.
As shown in Fig.~1(b), for the case of $\rho$=1,
when we decrease $\beta_4$ from zero,
first $\langle q \rangle$ is gradually decreased,
since the bottom of the on-center type potential
is not so widened by $\beta_4$.
Around at $\beta_4$$\approx$$-0.0025$, corresponding to
the region of the rattling type,
$\langle q \rangle$ is rapidly decreased.
When we further decrease $\beta_4$,
$\langle q \rangle$ is gradually decreased
for $\beta_4$$<$$-0.003$ in accordance with the shift of
potential minima.
For $\rho$=2, $\langle q \rangle$ is smoothly decreased
in the whole region of $\beta_4$$<$0.
Note that in the regions for on-center and rattling types,
we observe significant difference between the results for
$\rho$=1 and 2.
Since the electron-phonon coupling for $\rho$=2 is
virtually stronger than that for $\rho$=1,
the positions of potential minima for $\rho$=2 are
larger than that for $\rho$=1.
However, in the region of the off-center type,
the potential is heavily deformed by large negative value
of $\beta_4$ even without the electron-phonon coupling.
Thus, the position of the potential minima is insensitive
to the electron number for $\beta_4$$<$$-0.003$.


The difference in the dependence of the change of the potential
shape on electron number can be significantly found
in the effective attraction between electrons mediated by phonons.
When we define the attraction as $-U_{\rm ph}$,
the magnitude of the attraction $U_{\rm ph}$ is given by
\begin{equation}
  U_{\rm ph} = 2E^{(0)}_1-(E^{(0)}_0+E^{(0)}_2),
\end{equation}
where $E^{(0)}_{\rho}$ is the ground-state energy of
$H_{\rm eph}$ for the electron number $\rho$.
We note that the effect of $U$ is not included in $E^{(0)}_{\rho}$.
Note also that $U_{\rm ph}$=$2\alpha\omega$ at half-filling
for harmonic phonon.

In Fig.~1(c), we plot $U_{\rm ph}/\omega$ as a function of $\beta_4$.
At $\beta_4$=0, $U_{\rm ph}/\omega$ is about 3.7,
which is close to $2\alpha$ for $\alpha$=2.
When we decrease $\beta_4$, $U_{\rm ph}$ is increased,
since the degree of the potential deformation for $\rho$=2
is relatively larger than that for $\rho$=1.
When the bottom of the potential becomes flat and wide,
the phonon density of states at low energies are increased.
Then, the polaronic effect is enhanced,
leading to the increase of the attraction.
However, in the region of the off-center type,
there is no significant difference in
the degree of the potential deformation between
the cases for $\rho$=1 and 2, as understood from Fig.~1(b).
Then, $U_{\rm ph}$ is decreased when we further decrease
$\beta_4$ in the region of the off-center type potential.\cite{note}

From the above result, we can understand that the Coulomb
interaction is effectively reduced for the potential shape
of the rattling type.
For a reasonable choice of the parameter as $U/\omega$=10,
we actually find a region with effective attraction
around at $\beta_4$=$-0.0025$.
This suggests that the Kondo temperature should be enhanced
in such a region.
In the following, let us focus on the Kondo effect
for $-0.003$$<$$\beta_4$$<$$-0.002$.


Now we include the hybridization between localized and
conduction electrons.
Since we consider localized oscillation of the caged atom,
we ignore the coupling between phonons and conduction electrons,
Then, the Hamiltonian is given by
\begin{equation}
  \label{HAmodel}
  H \!=\! \sum_{\mib{k}\sigma} \varepsilon_{\mib{k}}
  c_{\mib{k}\sigma}^{\dag} c_{\mib{k}\sigma}
  \!+\!
  \sum_{\mib{k}\sigma} (Vc_{\mib{k}\sigma}^{\dag}d_{\sigma}+{\rm h.c.})
  \!+\! \varepsilon_{\rm d}\rho \!+\! H_{\rm loc},
\end{equation}
where $\varepsilon_{\mib{k}}$ denotes the dispersion of conduction electron,
$c_{\mib{k}\sigma}$ is an annihilation operator of conduction electron
with momentum $\mib{k}$ and spin $\sigma$,
$V$ is the hybridization between conduction and localized electrons,
and $\varepsilon_{\rm d}$ indicates a local electron level.
In the following, we always adjust $\varepsilon_{\rm d}$ to consider
the half-filling case, i.e., $\langle \rho \rangle$=1,
even though we do not mention explicitly.
The energy unit is chosen as a half of the conduction bandwidth,
which is expressed by $D$.
Hereafter we fix $V/D$=0.25 and $U/D$=2.
Typically, $D$ is in the order of 1 eV.

In order to clarify electronic properties of $H$,
we evaluate entropy $S_{\rm imp}$, specific heat $C_{\rm imp}$, and
susceptibility by using a numerical renormalization group (NRG) method.
\cite{NRG}
The logarithmic discretization of the momentum space
is characterized by a parameter $\Lambda$
and we keep $M$ low-energy states
for each renormalization step.
Here we set $\Lambda$=2.5 and $M$=5000.
Spin and charge susceptibilities are, respectively,
given by
\begin{equation}
  \begin{split}
  & \chi_{{\rm s}} =
  \frac{1}{Z} \sum_{n,m}
  \frac{e^{-E_n/T}-e^{-E_m/T}}{E_m-E_n}
  |\langle n | \sigma_{z} | m \rangle|^2,\\
  & \chi_{{\rm c}} =
  \frac{1}{Z} \sum_{n,m}
  \frac{e^{-E_n/T}-e^{-E_m/T}}{E_m-E_n}
  |\langle n | (\rho- \langle \rho \rangle) | m \rangle|^2,
  \end{split}
\end{equation}
where $E_n$ is the eigenenergy for the $n$-th eigenstate
$|n\rangle$ of $H$,
$T$ is a logarithmic temperature defined as $T$=$D\Lambda^{-(N-1)/2}$,
$Z$ is the partition function given by $Z$=$\sum_n e^{-E_n/T}$,
$\langle \rho \rangle$=
$(1/Z)\sum_{n} e^{-E_n/T} \langle n | \rho | n \rangle$,
and $\sigma_z$=$n_{\uparrow}$$-$$n_{\downarrow}$.

\begin{figure}[t]
\label{fig2}
\centering
\includegraphics[width=6.1truecm]{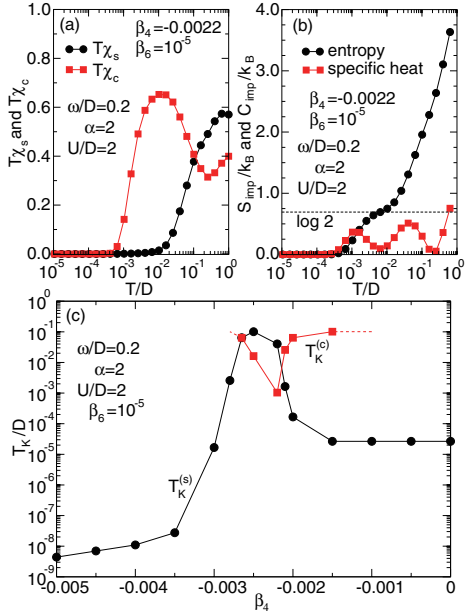}
\caption{(Color online)
(a) $T\chi_{\rm s}$ and $T\chi_{\rm c}$ vs. temperature.
(b) $S_{\rm imp}$ and $C_{\rm imp}$ vs. temperature.
(c) Kondo temperatures vs. $\beta_4$.
}
\end{figure}


In Figs.~2(a) and 2(b), we show typical results for
susceptibility, entropy, and specific heat
for $\beta_4$=$-0.0022$.
Note that this is the region with weak effective attraction,
given by $U$$-$$U_{\rm ph}$=$-0.075D$.
Thus, the spin susceptibility is rapidly suppressed due to
the effect of on-site attraction, while the
charge susceptibility is increased without the slight decrease
near $T/D$=1.
When the spin susceptibility is suppressed,
spin entropy $\log 2$ is released,
leading to a peak in the specific heat.
This peak temperature is defined as $T_{\rm K}^{\rm (s)}$.

After the release of the spin entropy, we observe
a narrow plateau of $\log 2$, corresponding to
the degenerate region characterized by vacant and double
occupied states.
This entropy of $\log 2$ is gradually released and
it eventually goes to zero around at $T/D$$\sim$$10^{-3}$,
where we can see another peak in the specific heat.
We define this temperature as $T_{\rm K}^{\rm (c)}$.

Changing the values of $\beta_4$, we repeat the calculations
and plot the Kondo temperatures in Fig.~2(c).
As understood from Fig.~2(b), we can observe a couple of peaks
in the specific heat, corresponding to the vanishment of
spin and charge susceptibilities.
Thus, we carefully check the behavior of $T\chi_{\rm s}$ and
$T\chi_{\rm c}$,
in order to specify the origin of the entropy release,
leading to the two kinds of the Kondo temperature.

In the region of the on-center type potential,
$T_{\rm K}^{\rm (s)}$ is not so affected by $\beta_4$
and we cannot find
a high-temperature peak
corresponding to
$T_{\rm K}^{\rm (c)}$.
However, when we approach the region of the rattling-type
potential,
$T_{\rm K}^{\rm (s)}$ is increased and $T_{\rm K}^{\rm (c)}$
appears due to the reduction of the Coulomb interaction.
Eventually, the effective interaction turns out to be attractive
and $T_{\rm K}^{\rm (c)}$ becomes lower than $T_{\rm K}^{\rm (s)}$.
When we further decrease $\beta_4$, we enter the region of
the off-center type and the phonon-mediated attraction
is rapidly decreased, as found in Fig.~1(c).
The peak corresponding to the entropy release of charge
degree of freedom is changed to just a shoulder-like structure,
and it finally disappears.

In general, when the Kondo temperature is increased,
we can expect that $\gamma$ is magnetically robust.
Roughly speaking, the electronic properties will not be
significantly affected by a magnetic field $H$ smaller than
$k_{\rm B} T_{\rm K} /\mu_{\rm B}$,
where $\mu_{\rm B}$ is a Bohr magneton.
Thus, in the region of the rattling-type potential
with enhanced $T_{\rm K}$,
we can expect that $\gamma$ is magnetically robust.
On the other hand,
in the Kondo problem, it has been known that $\gamma$ is
in proportion to $1/T_{\rm K}$.\cite{review}
As we will describe later, the behavior of $\gamma$
as a function of
$\beta_4$ is well explained by that of $1/T_{\rm K}$.
When the Coulomb repulsion is weakened by the phonon-mediated
attraction, the Kondo temperature is increased and thus,
$\gamma$ is decreased.
Namely, the magnitude of $\gamma$ is not expected to be
large in the region of the rattling type.

\begin{figure}[t]
\label{fig3}
\centering
\includegraphics[width=6truecm]{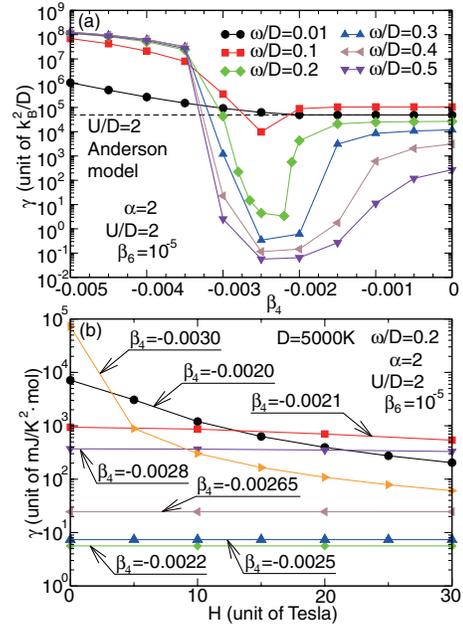}
\caption{(Color online)
(a) Sommerfeld constant $\gamma$ vs. $\beta_4$ for several
values of $\omega$.
(b) $\gamma$ in the unit of mJ/K$^2$$\cdot$mol vs.
magnetic field for $D$=5000 K.
}
\end{figure}


However, in the simple discussion based on the relation
of $\gamma$$\sim$$1/T_{\rm K}$,
it is unclear whether the magnitude of $\gamma$ is large or not.
Thus, we actually evaluate $\gamma$, given by
$\gamma$=$C_{\rm imp}/T$ at low temperatures.
Here $\gamma$ is estimated at the lowest temperature
at which we can reach in the present NRG calculation.

In Fig.~3(a), we show $\gamma$ vs. $\beta_4$
for several values of $\omega$.
For $\omega/D$=0.01, in which the adiabatic approximation
is considered to be valid,
the value of $\gamma$ is not so sensitive to $\beta_4$,
even if the potential shape is changed by $\beta_4$.
In particular, for $\beta_4$$>$$-0.002$, the values of $\gamma$
are similar to that of the Anderson model without phonons.
The deviation can be found only in the region of
the potential of the off-center type.

When we increase the value of $\omega$, the effect of
anti-adiabaticity appears.
For $\omega$=0.2, $\gamma$ decreases significantly
for $-0.003$$<$$\beta_4$$<$$-0.002$,
which is, roughly speaking, corresponding to 
the region of the rattling-type potential.
Note that this behavior of $\gamma$ is well reproduced
by $1/T_{\rm K}^{\rm (c)}$.
When we further increase $\omega$, the region of
the suppressed $\gamma$ becomes wide and
for $\omega/D$=$0.5$, such a region appears for
$\beta_4$$>$$-0.003$.
Note that very large values of $\gamma$ can be found only
in the region of the off-center type potential.

Now we turn our attention to the effect of a magnetic field.
In Fig.~3(b), we show $\gamma$ vs. magnetic field $H$
along the $z$-axis for $-0.003$$\le$$\beta_4$$\le$$-0.002$
with $\alpha$=2, $\omega/D$=$0.2$, and $\beta_6$=$10^{-5}$.
In this calculation, we add the Zeeman term
$H_{\rm mag}$=$\mu_{\rm B} \sigma_z H$
to the Hamiltonian (\ref{HAmodel}),
where the $g$-factor is set as 2.
To evaluate the values of $\gamma$ in the unit of
mJ/K$^2$$\cdot$mol, we set $D$=5000 K.
For $\beta_4$ deviated from the rattling-type potential,
the magnitude of $\gamma$ at $H$=0 is enhanced,
but it decreases with the increase of $H$,
as found in the curves for $\beta_4$=$-0.002$ and $-0.003$.
On the other hand, when we take $\beta_4$
corresponding to the rattling-type potential,
$\gamma$ is actually magnetically robust, but
the magnitude of $\gamma$ is not so large.
However, when we consider the parameter region just between
the on-center and the rattling types ($\beta_4$=$-0.0021$)
or between the off-center and the rattling types ($\beta_4$=$-0.0028$),
it seems to be possible to reproduce magnetically robust
$\gamma$ in the order of 100 mJ/K$^2$$\cdot$mol.
These regions are corresponding to the cases at which
the Coulomb repulsion is canceled by phonon-mediated attraction.
Thus, on the basis of a concept of the Kondo effect in
the electron-phonon system,
we propose that magnetically robust heavy fermion behavior is
understood by the competition between spin and charge fluctuations
at the region of $T_{\rm K}^{\rm (s)}$$\sim$$T_{\rm K}^{\rm (c)}$.


Here we provide a couple of comments.
(i) It has been pointed out that
the competition between the Coulomb interaction and phonon-mediated
attraction plays an important role in strongly coupled
electron-phonon system.
\cite{Freericks3,Hotta-Takada1a,Hotta-Takada1b,Hotta-Takada2,Hotta-Takada3,Koller}
Here we control such a competition by anharmonicity
to explain magnetically robust heavy fermion phenomenon.
(ii) We have considered the one-dimensional potential for simplicity,
but in actuality, the motion of the caged atom occurs
in a three-dimensional potential.
In particular, there exists a mode in which
the caged atom moves around potential minima.\cite{Takabatake}
Concerning this issue, the present author has analyzed the Anderson model
coupled with Jahn-Teller phonons.\cite{Hotta1}
Then, we have proposed an intriguing phenomenon of chiral Kondo effect
due to the rotational degree of freedom of Jahn-Teller phonon.\cite{Hotta2}
On the basis of this effect, the magnetically robust heavy fermion
phenomenon may be explained for the off-center type potential.
It is one of future issues.

Finally, we briefly discuss the effect of orbital degree of freedom,
which has been ignored in this paper.
Although the motivation of this work is
magnetically robust heavy fermion phenomenon
observed in Sm-based filled skutterudites,
we have not considered $f$ electrons in the model.
In $f$-electron systems with orbital degeneracy, multipole degree of freedom
becomes active due to strong spin-orbit coupling.
In this sense, in order to discuss electronic properties of
actual Sm-based filled skutterudites,
it is important to consider multipole degrees of freedom.\cite{Hotta4}
The multi-orbital Anderson model coupled with Jahn-Teller phonons
has been also analyzed,\cite{Hotta5}
but further works are required along with this research direction,
in particular, in order to understand the difference in
SmOs$_4$Sb$_{12}$ and SmFe$_4$P$_{12}$.
It is also expected that more stable magnetically robust heavy fermion
state is realized when orbital fluctuations are also comparable with
spin and charge fluctuations.
It is another future problem.


In summary, we have discussed the Kondo effect of
the Anderson model coupled with local anharmonic phonon.
In order to understand the magnetically robust heavy fermion phenomenon,
we have emphasized a key issue of the shape of the cage potential.
In particular, we have found that $\gamma$ is
both large and magnetically robust,
when the potential shape is deformed from the rattling type,
in which spin and charge fluctuations are comparable to each other.


The author is grateful to Y. Aoki and H. Sato for
fruitful discussions and useful comments.
He also thanks R. Miyazaki for his contribution to
the calculation of local phonon-mediated attraction.
This work has been supported by a Grant-in-Aid for
Scientific Research (C)
from Japan Society for the Promotion of Science.
The computation in this work has been done using the facilities
of the Supercomputer Center of Institute for Solid State Physics,
University of Tokyo.



\begin{thebibliography}{99}
\bibitem{note}
Note that $U_{\rm ph}$ shows an abrupt change
at $\beta_4$=$-\sqrt{\beta_6}$=$-0.00316$,
when three potential minima take the same value.
Such an abrupt change should be smeared when electrons
are delocalized.
\end{thebibliography}
\end{document}